\documentclass[%
 aip,
 amsmath,amssymb,
 reprint,%
]{revtex4-1}

\usepackage{graphicx}
\usepackage{dcolumn}
\usepackage{bm}
\usepackage{cancel}

\usepackage[utf8]{inputenc}
\usepackage[T1]{fontenc}
\usepackage{mathptmx}
\usepackage{etoolbox}

\makeatletter
\def\@email#1#2{%
 \endgroup
 \patchcmd{\titleblock@produce}
  {\frontmatter@RRAPformat}
  {\frontmatter@RRAPformat{\produce@RRAP{*#1\href{mailto:#2}{#2}}}\frontmatter@RRAPformat}
  {}{}
}%
\makeatother
\begin{document}
\preprint{AIP/123-QED}

\title{Global efficiency and network structure of urban traffic flows: A percolation-based empirical analysis}
\author{Yungi Kwon}
\affiliation{Natural Science Research Institute, University of Seoul, Seoul 02504, Republic of Korea}

\author{Jung-Hoon Jung}
\affiliation{Department of Physics, University of Seoul, Seoul 02504, Republic of Korea}

\author{Young-Ho Eom}
\email{yheom@uos.ac.kr}
\affiliation{Natural Science Research Institute, University of Seoul, Seoul 02504, Republic of Korea}
\affiliation{Department of Physics, University of Seoul, Seoul 02504, Republic of Korea}
\affiliation{Urban Big data and AI Institute, University of Seoul, Seoul 02504, Republic of Korea}

\date{\today}

\begin{abstract}
Making the connection between the function and structure of networked systems is one of the fundamental issues in complex systems and network science. Urban traffic flows are related to various problems in cities and can be represented as a network of local traffic flows. To identify an empirical relation between the function and network structure of urban traffic flows, we construct a time-varying traffic flow network of a megacity, Seoul, and analyze its global efficiency with a percolation-based approach. Comparing the real-world traffic flow network with its corresponding null-model network having a randomized structure, we show that the real-world network is less efficient than its null-model network during rush hour, yet more efficient during non-rush hour. We observe that in the real-world network, links with the highest betweenness tend to have lower quality during rush hour compared to links with lower betweenness, but higher quality during non-rush hour. Since the top betweenness links tend to be the bridges that connect the network together, their congestion has a stronger impact on the network's global efficiency. Our results suggest that the spatial structure of traffic flow networks is important to understand their function.
\end{abstract}
\maketitle

\begin{quotation}
Traffic flows in a city can be represented as a network of local traffic flows on individual roads. How do efficient and inefficient traffic flow networks differ in structure? To answer this question, we compare a time-varying real-world traffic flow network to networks with a randomized structure. We find that the real-world network is less efficient during rush hour, but more efficient at other times than the randomized networks. We show that links that span the real-world network are more prone to congestion during rush hour than links that do not, while being faster at other times. Because the spanning links keep the network's global connectivity, the observed difference in the efficiency occurs. Our findings indicate that such spanning links can be an effective target to detect urban traffic congestion, and the relationship between the network structure and function of urban traffic flows helps one to understand urban traffic congestion.
\end{quotation}

\section{Introduction}\label{sec:intro} 
A fundamental issue in complex systems and network science is understanding the relationship between the structure and function of networked systems~\cite{Newman2018Networks,Barabasi2016Network, holovatch2017complex}. It is well known, for example, that the degree distribution of a given network can tell us about how epidemics spread out in the network~\cite{pastor2001epidemic, pastor2015epidemic} and how robust the network is to node or link removal~\cite{Albert2000Error,Callaway2000NetRbustness}. 

Urban traffic flows are an emergent phenomenon relevant to various issues such as congestion~\cite{de2009congestion, zhao2005onset, tan2014traffic, sole2016congestion, ccolak2016understanding, olmos2018macroscopic}, pollution~\cite{rosenlund2008comparison, zhang2013air}, and well-being~\cite{weisbrod2003measuring, sweet2014traffic, hennessy1999traffic} in cities. An effective way to understand the traffic flows of a city is to represent them as a network of local traffic flows between intersections on the underlying road network, the conduits for those flows. In this traffic flow network, each link is characterized by the quality of the local traffic flow corresponding to that link. Like other studies~\cite{li2015percolation, zeng2019switch,zeng2020multiple,ruan2019empirical,zhang2017comparison}, we use a relative velocity of a flow as its quality. See Sec.~\ref{subsec:dataset} for the details. Therefore, the structure of a traffic flow network is, in general, different from that of its underlying road network. We are interested in what structure real-world traffic flow networks have~\cite{hamedmoghadam2017complex, taillanter2021empirical}, how the structure of the networks changes over time, and the relationship between the structure and function of the networks~\cite{berezin2015localized, hamedmoghadam2022percolation, gao2022resilient,louf2013modeling,louf2014congestion}. 

A percolation-based approach~\cite{li2015percolation, zeng2019switch, zeng2020multiple, hamedmoghadam2021percolation,cogoni2021stability,ruan2019empirical,zhang2017comparison} to traffic flow networks is recently attracting attention as it quantifies the global efficiency of the networks based on their structural properties. The basic idea is that how quickly vehicles can reach most of a given traffic flow network represents its global efficiency. In this approach, a given traffic flow network fragments into parts when one removes all the links whose quality is lower than or equal to the percolation threshold from the network. The percolation threshold indicates the maximum quality that one can travel over the main part of the network and, thus, can be interpreted as a proxy for the global efficiency of the network. This approach identified bottleneck links~\cite{li2015percolation,hamedmoghadam2021percolation}, stable breakup patterns at criticality~\cite{cogoni2021stability}, the impacts of weather on urban traffic~\cite{ruan2019empirical}, and multiple functional states in traffic flow networks~\cite{zeng2020multiple}, to name a few. A recent work~\cite{zeng2019switch} also reported that the value of a percolation critical exponent obtained from real-world traffic flow networks in rush hour is similar to the exponent observed in two-dimensional lattices, whereas in non-rush hour similar to the exponent observed in small-world networks. 

\begin{figure*}[!htb]
    \includegraphics[width=0.92\textwidth]{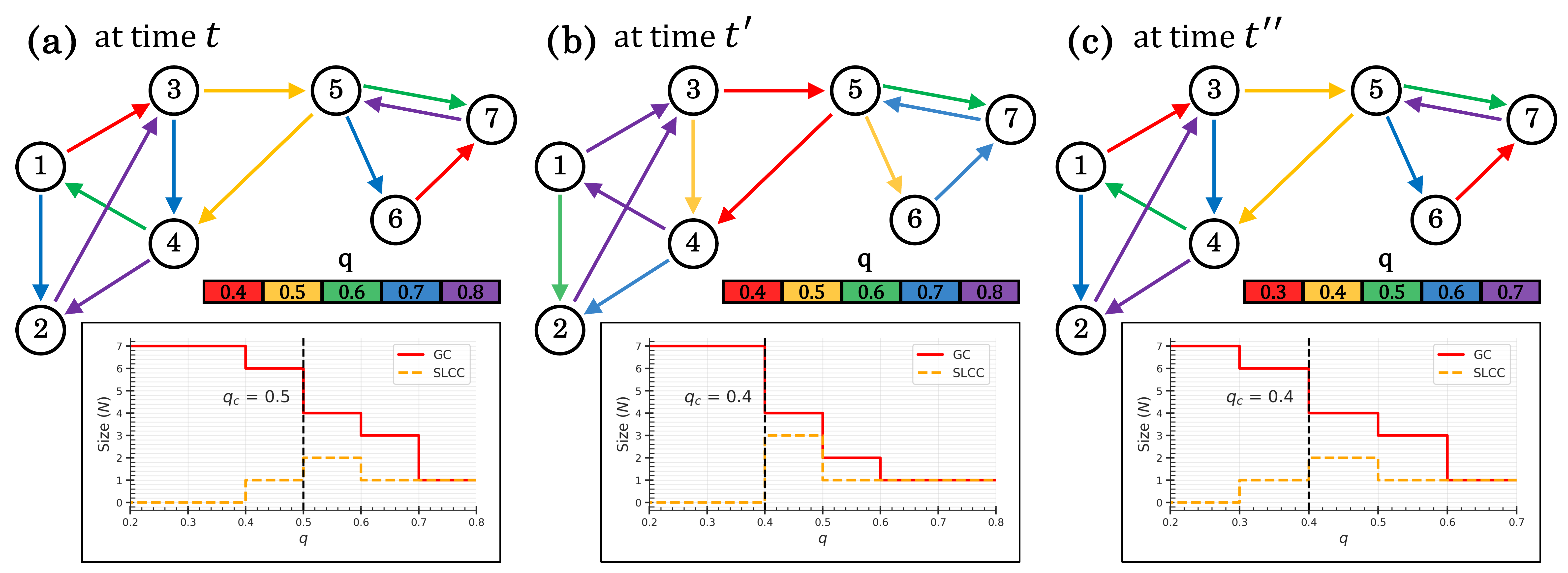}
    \caption{\textbf{Percolation process on toy traffic flow networks.} \textbf{(a)} A traffic flow network with $N=7$ nodes and $L=12$ links at time $t$, where each link has its quality, $q$, following the color-coded scheme right below. The percolation threshold, $q_c$, is determined to be 0.5 at which the network is disconnected and the size of the second-largest connected component is maximal. \textbf{(b)} The network at time $t'$ has the same $q$ distribution with (a), but a different spatial structure. The network is disconnected when we remove all the links with $q\leq0.4$ and thus the threshold $q_c$ is to be 0.4. \textbf{(c)} The network at time $t''$ in which the qualities of all links in (a) are decreased by 0.1 merely. As in (b), $q_c=0.4$.}
    \label{fig:fig1}
\end{figure*}

However, the existing percolation approach has a resolution limit to reveal the role of the network structure in the global efficiency of traffic flow networks because their percolation threshold, the proxy for their global efficiency, can be determined by not only their spatial structure  but also their link quality distribution. Let us consider, for instance, a hypothetical time-varying traffic flow network depicted in Fig.~\ref{fig:fig1} whose panels (a), (b), and (c) represent the traffic flow network at time $t$, $t'$, and $t''$, respectively. The percolation threshold, $q_c$, is 0.5 for the network at time $t$ (Fig.~\ref{fig:fig1}(a)). In other words, if we remove the links whose quality is lower than or equal to $0.5$, the network is disconnected. However, $q_c=0.4$ for the network at time $t'$ (Fig.~\ref{fig:fig1}(b)) although the network has the identical link quality distribution at time $t$ and $t'$. On the other hand, $q_c=0.4$ for the network at time $t''$, although the network has the same spatial structure at time $t$ and $t''$ in the sense that links with higher (lower) quality at time $t$ have higher (lower) quality at time $t''$ as shown in Figs.~\ref{fig:fig1} (a) and (c). Note that each link in the network has a quality difference of 0.1 at time $t$ and $t''$. These examples show that the global efficiency of a traffic flow network depends on two structural features: its spatial structure and link quality distribution and that we need to separate these features for a deeper understanding. In particular, the role of the spatial structure is less clear than the role of the link quality distribution.

In this article, we compare a real-world traffic flow network and its null-model network with the same link quality distribution, yet with a randomized spatial structure, in order to identify an empirical relation between the global efficiency and the spatial structure of urban traffic flow networks. We observed that the real-world network is less efficient than its null-model network during rush hour, but more efficient during non-rush hour. This indicates that the spatial structure of urban traffic flow networks during rush hour is qualitatively different from the spatial structure during non-rush hour. We identified that links with the highest betweenness~\cite{freeman1977set, girvan2002, newman2012} tend to have lower quality than links with low betweenness during rush hour, whereas higher quality during non-rush hour. Since the top betweenness links tend to be the bridges that connect the network together, their congestion has a stronger impact on the network efficiency. Our results suggest that real urban traffic congestion might arise when such top betweenness links are severely congested rather than the whole system is slowing down, and the relationship between the structure and function of traffic flow networks helps to understand urban traffic congestion.

\section{Dataset and methods}\label{sec:dataset_method}
\begin{figure*}[!ht]
    \includegraphics[width=\textwidth]{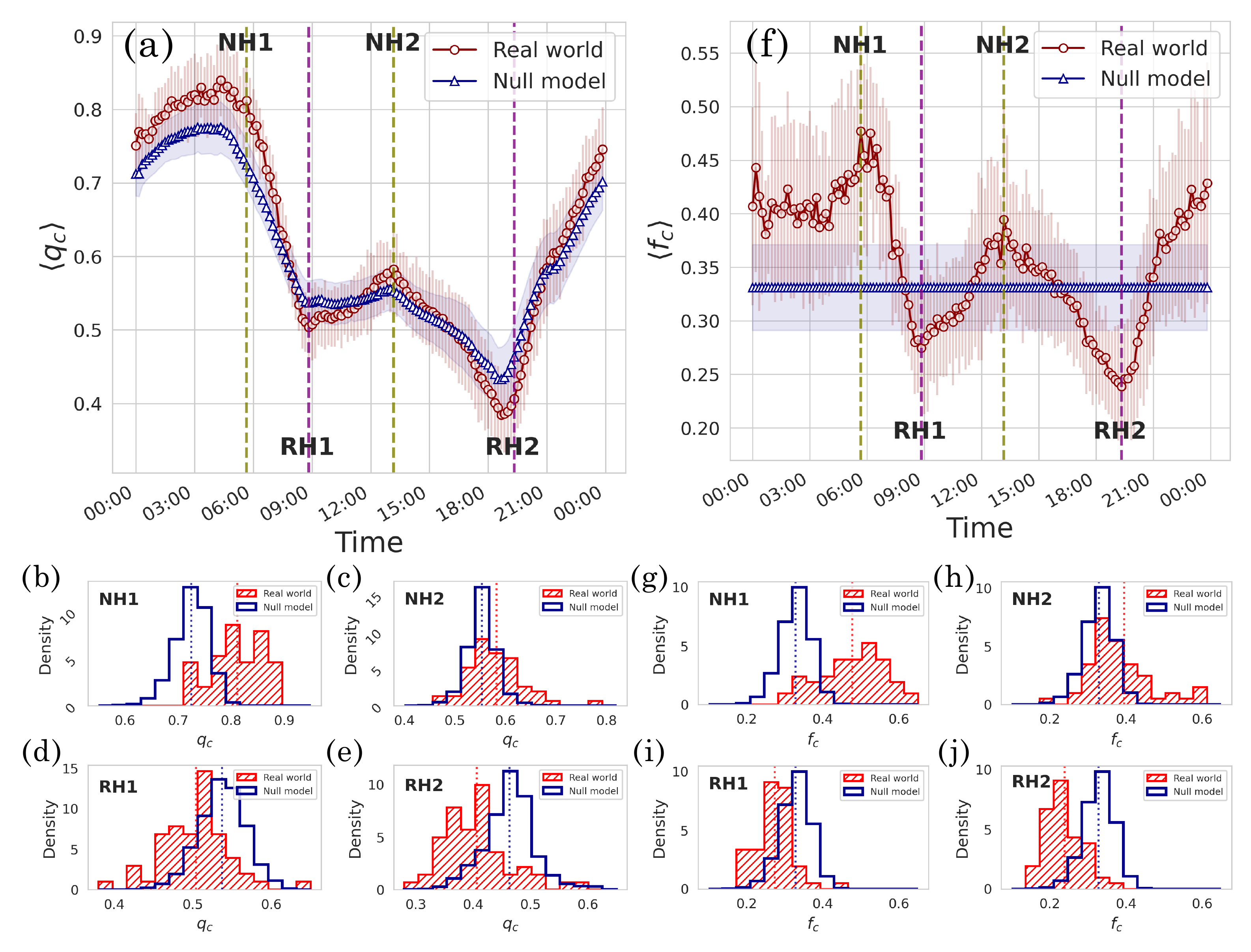}
    \caption{\textbf{Percolation threshold ($q_c$) and critical removal fraction ($f_c$) of the real-world and null-model traffic flow networks.} \emph{(a)} Average percolation threshold $\langle q_c(t) \rangle$ of the real-world traffic flow network and its null-model network. For the real-world case, $\langle q_c(t) \rangle$ is obtained by averaging over 57 workdays, and the error-bars for each instant denote the standard deviations (SDs) over the workdays. In the null-model cases, $\langle q_c(t) \rangle$ is obtained by averaging over 50,000 randomly shuffled realizations for each instant and the shaded region shows the SD over all realizations. Here we choose four representative times for non-rush hour (\textbf{NH1(05:40), NH2(13:10)}) and rush hour (\textbf{RH1(08:50), RH2(19:20)}) with the greatest difference in $\langle q_c(t) \rangle$ between the real-world and null-model networks, denoted by dashed lines with olive and purple colors, respectively. \emph{(b)-(e)} The distribution of $q_c$, for all workdays, measured at \textbf{NH1, NH2, RH1}, and \textbf{RH2}, where each dotted line denotes the mean of the distribution. (f) Average critical fraction $\langle f_c(t) \rangle$ of the real-world network and its null-model networks. \emph{(g)-(j)} The distribution of $f_c$, for all workdays, measured at \textbf{NH1, NH2, RH1}, and \textbf{RH2}, where each dotted line denotes the mean of the distribution.}
    \label{fig:fig2}
\end{figure*} 

\subsection{Construction of the Seoul traffic flow network }\label{subsec:dataset}
First, we built a directed network based on the Seoul's road network, which consists of 1756 intersections and 4924 road segments. These intersections and road segments correspond to the nodes and the links of our traffic flow network, respectively. We weighted each link with the functional quality of the traffic flowing over it based on the real-time velocity data of the Seoul road traffic, which is recorded from December 1, 2020 to February 28, 2021 (57 workdays and 30 days off), with a resolution of 5 min. We adopted the link quality quantification method from previous studies~\cite{li2015percolation,zeng2019switch} as described below. For a directed link from node $i$ to $j$, we first identified the 95th percentile of its velocity $v_{ij}^{95th}$ in each day as its effective maximal flow performance. Then, based on this velocity, we normalized each flow's instant velocities at time $t$ in the same day and defined the flow quality of the link, $q_{ij}(t)$, as $q_{ij}(t) = v_{ij}(t) / v_{ij}^{95th}$. In this way, we built a series of weighted networks of local traffic flows at every recorded instant for analysis. Note that all analyses conducted in our research were performed separately on workdays and days off. Unless otherwise stated, all results presented in this paper are for workdays.

\subsection{Percolation analysis and null-model networks }\label{subsec:method}
Adopted from previous studies~\cite{li2015percolation,zeng2019switch}, our percolation-based analysis of traffic flow networks was conducted as follows. For a given threshold $q$, we remove all links whose qualities are equal to or lower than the threshold. Starting from $0$, we continuously increase the value of the threshold and keep track of the sizes of the giant component (GC), the largest connected component in our case, and the second-largest connected component (SLCC). Note that both the GC and the SLCC are strongly connected components in this study. The percolation threshold $q_c$ is determined when the SLCC is at its maximum~\cite{li2015percolation, hamedmoghadam2021percolation, wang2022percolation}. At the percolation threshold, the size of the GC is expected to drop the most, but the size of the SLCC typically leads to a refined and more evident peak~\cite{hebert-dufresne2019smeared}. Some example cases of percolation analysis are illustrated in the insets of Fig.~\ref{fig:fig1}. The percolation threshold $q_c$ of a given network is the maximum relative velocity that one can traverse most parts of the network, and can, thus, be interpreted as a proxy for its global efficiency, because the reachability and accessibility of any transportation network are the minimum requirements for its function~\cite{Albert2000Error,Callaway2000NetRbustness,li2021percolation}. 

To reveal the role of the spatial structure of a real-world traffic flow network in its global efficiency, we need to compare it with a null-model network with the same link quality distribution but with a randomized spatial structure as discussed in the Sec.~\ref{sec:intro}. A simple way to create such a null-model network is to randomly shuffle the link qualities in the real-world network since the shuffling does not change the link quality distribution but makes the spatial structure of the networks randomized. In this study, we use such a network with randomly shuffled weights as the null model, leaving the node and link information intact from the real-world networks.

\begin{figure*}[!ht]
     \includegraphics[width=\textwidth]{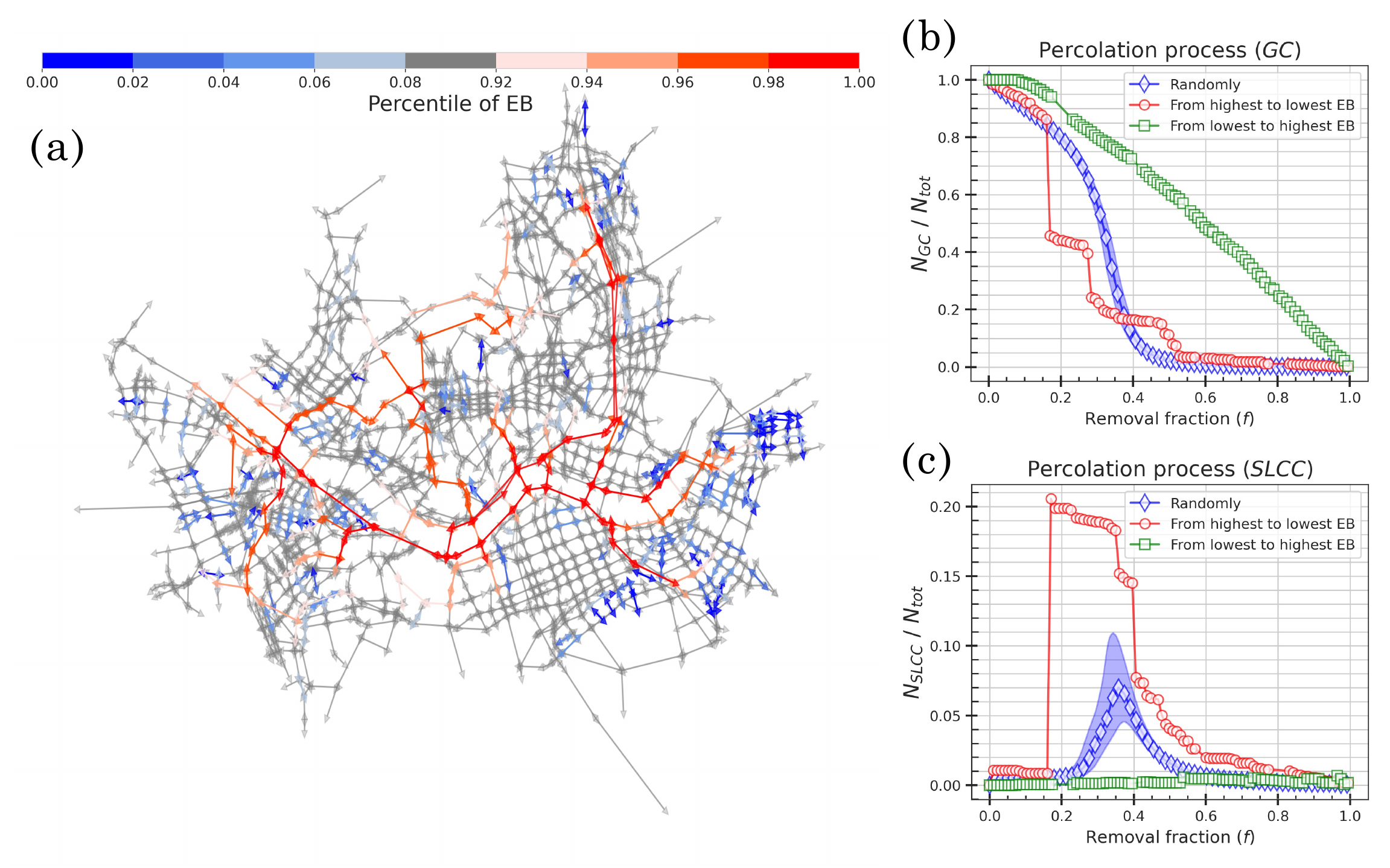}
     \caption{\textbf{Edge betweenness and its impact on the percolation process.} \textbf{(a)} Edge betweenness distribution on the Seoul traffic flow network. \textbf{(b)} Normalized size of $GC$ as a function of the removal fraction, $f$, of links in different percolation processes. \textbf{(c)} Normalized size of $SLCC$ as a function of $f$ of links in different percolation processes. The random percolation processes, depicted in (b) and (c), are derived by 50000 realizations. For each $f$, the marker and the shaded area denote the median and the range for the 25th-75th percentile of the distribution, respectively.
}         
     \label{fig:fig3}
\end{figure*}

\section{Results}\label{sec:case}
We obtain the percolation threshold $q_c(t)$ at a given time $t$, which quantifies the global efficiency of the Seoul traffic flow network at that time, by removing the links from the network based on their quality. The detailed procedure of the percolation process is described in Sec.~\ref{subsec:method}. Fig.~\ref{fig:fig2}(a) shows the intra-day fluctuations in the average percolation threshold, $\langle q_c(t) \rangle$, of the Seoul traffic flow network (empty circle symbols). The average value $\langle q_c(t) \rangle$ for a given time $t$ is obtained by averaging the percolation thresholds of 57 workdays. Roughly speaking, the value of $\langle q_c(t) \rangle$ is very high from 0:00 a.m. to 6:00 a.m., but it decreases rapidly from 6:00 a.m. to 9:00 a.m. It slightly recovers from 9:00 a.m. to 1:00 p.m., declines from 1:00 p.m. to 7:00 p.m., and rapidly recovers after 7:00 p.m. Similar patterns were observed in Beijing, although the values of $q_c(t)$ in the morning rush hour tend to be lower than the values in the evening rush hour in the case of Beijing~\cite{li2015percolation}. However, it is not clear what makes the observed fluctuating patterns in the global efficiency of the real-world traffic flow networks in Seoul and Beijing.

To identify the effect of the spatial structure of traffic flow networks on their global efficiency, we compare the Seoul traffic flow network with its corresponding null-model network with the same link quality distribution but with a randomly shuffled structure at each time. The detailed procedure for generating the null-model network is described in Sec.~\ref{subsec:method}. As depicted in Fig.~\ref{fig:fig2}(a), the average critical threshold $\langle q_c(t) \rangle$ of the Seoul traffic flow network is lower than that of its null model during rush hour (e.g., from 8:00 a.m. to 11:25 a.m. and from 4:10 p.m. to 8:45 p.m.), whereas higher during non-rush hour (e.g., from 11:25 a.m. to 4:10 p.m. and from 8:45 p.m. to 8:00 a.m.). Due to the large standard deviation in the average thresholds, we estimate the distribution of the percolation threshold $q_c(t)$ at 5:40 a.m. (Fig.~\ref{fig:fig2}(b)), 1:10 p.m. (Fig.~\ref{fig:fig2}(c)), 8:50 a.m. (Fig.~\ref{fig:fig2}(d)), and 7:20 p.m. (Fig.~\ref{fig:fig2}(e)). These distributions support the observation that the Seoul traffic flow network tends to be globally less efficient than its null-model network during rush hour, but more efficient during non-rush hour, indicating that there are qualitatively different efficiency states in the Seoul traffic flow network depending on its spatial structure.

\begin{figure*}[!ht]
     \includegraphics[width=\textwidth]{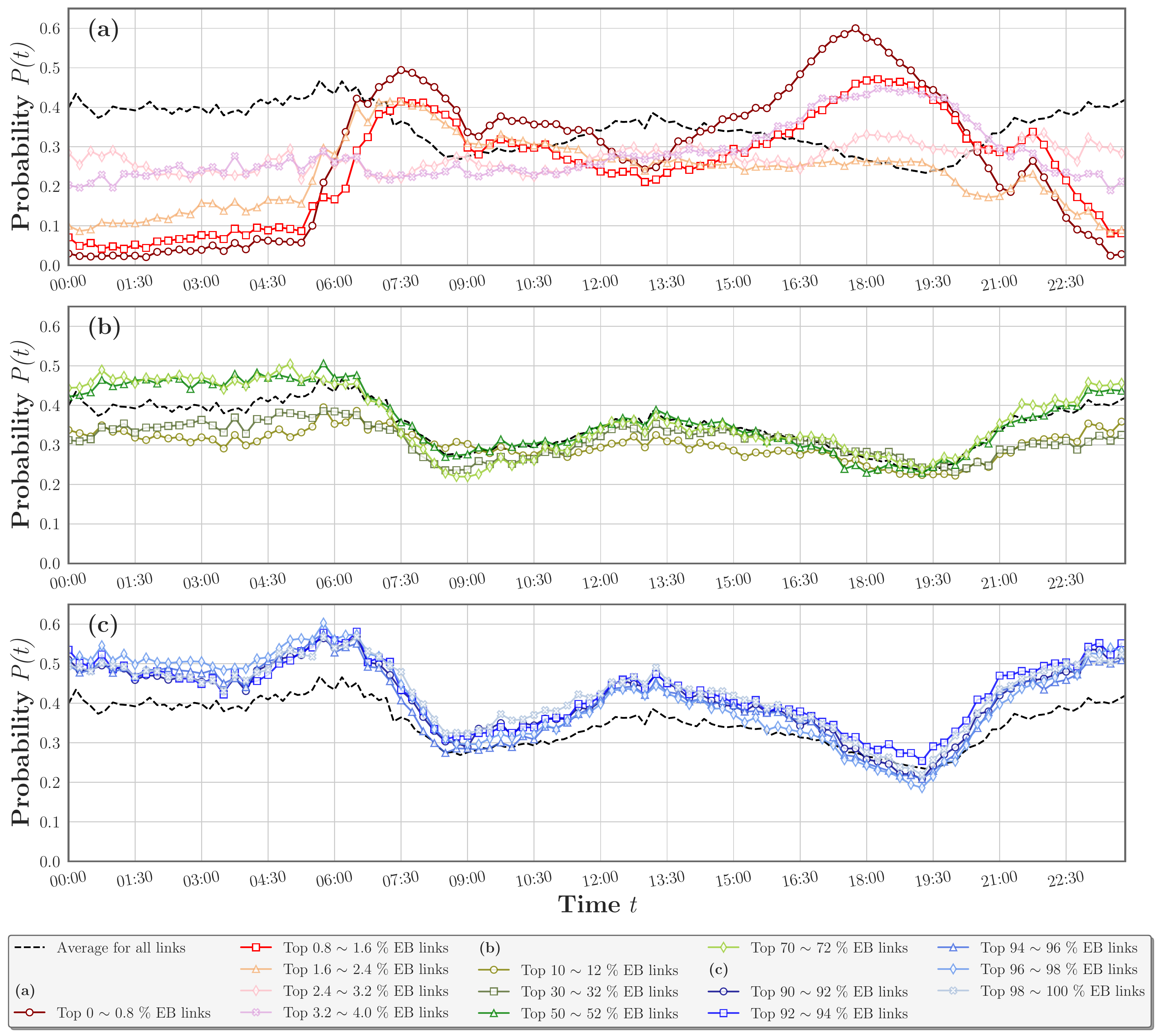}
     \caption{\textbf{Probability $P(t)$ that a link possesses quality lower than the percolation threshold $q_c(t)$.} \textbf{(a)} The average probabilities by groups, divided into 0.8\% increments from 0\% to 4\%. \textbf{(b)} The average probabilities by groups not belonging to the top and bottom EB. \textbf{(c)} The average probabilities by the bottom EB groups.}
     \label{fig:fig4}
\end{figure*}

To address the difference in the spatial structure of the real-world networks during rush hour and non-rush hour, first we check the critical removal fraction, $f_c(t)$, which represents the fraction of removed links when the percolation process based on link quality reaches the percolation threshold at time $t$. In other words, the critical removal fraction $f_c(t)$ is the ratio of the links in the network whose qualities are equal to or lower than $q_c(t)$. The time series of the average critical removal fraction, $\langle f_c(t) \rangle$, is depicted in Fig.~\ref{fig:fig2}(f). Note that $\langle f_c(t) \rangle$ of the null-model network appears to be constant because the percolation process is equivalent to random percolation as the link quality is randomly distributed in the network. The average removal fraction $\langle f_c(t) \rangle$ of the Seoul traffic flow network is lower than that of its null model during rush hour, while higher during non-rush hour. We also estimate the distribution of the critical removal fraction $f_c(t)$ at 5:40 a.m. (Fig.~\ref{fig:fig2}(g)) and 1:10 p.m. (Fig.~\ref{fig:fig2}(h)), 8:50 a.m. (Fig.~\ref{fig:fig2}(i)), and 7:20 p.m. (Fig.~\ref{fig:fig2}(j)). These results show that, when we remove low quality links from the real-world network, the smaller fraction of link removal is sufficient to break the network compared to the null model during rush hour, but the larger fraction of link removal is necessary during non-rush hour. This indicates that the quality of links that are important to keep the real-world network connected tends to be lower than the quality of other links during rush hour, but higher during non-rush hour.

Then which links are more important for the global connectivity of a given network? A possible answer would be links with high edge betweenness (EB)~\cite{akbarzadeh2019role,akbarzadeh2018communicability}. Since the EB of a link quantifies the number of shortest paths that run along the edge~\cite{girvan2002}, links with high EB of a given network are expected to act as bridges that connect the entire network. Links with EB in the Seoul traffic flow network are depicted in Fig.~\ref{fig:fig3}(a). Note that we obtained the EB of each link based on the unweighted version of the Seoul traffic flow network, whose topology is equivalent to the topology of the underlying road network. High EB links tend to be located at the center of the network and have a structure spanning the entire network, whereas low EB links are clustered, localized, and located at the periphery of the network. Hence we can hypothesize that these high EB links are important for keeping the whole network connected. To verify the above hypothesis, we performed a numerical experiment, which removes links from the Seoul traffic flow network in three specific orders: (i) ascending order of EB (i.e., from the lowest to the highest),  (ii) random order, and (iii) descending order of EB (i.e., from the highest to the lowest). Figs.~\ref{fig:fig3} (b) and (c) show that, when we remove links with descending order of EB, the critical removal fraction is $f_c=0.164$, which is much lower than the critical removal fraction of $f_c=0.33$ in the case of random link removal. On the other hand, if we remove links with ascending order of EB, the critical fraction is estimated to be a very high value of $f_c=0.958$, indicating that there is virtually no dramatic fragmentation of the GC. The numerical experiment indicates that links with high (low) EB are more (less) important to keep the real-world network connected.

To check whether the highest EB links are more prone to congestion during rush hour than other links, we measure the probability $P(t)$ for each link that the link’s quality is below the critical percolation threshold $q_c(t)$ among 57 working days (and thus, the link is removed before the network breaks during the percolation process) at each time $t$. Fig.~\ref{fig:fig4} shows the obtained probabilities averaged over links with a specific range of EB during the day. As shown in Fig.~\ref{fig:fig4}(a), top $0-1.6\%$ EB links are much more likely to have lower quality than $q_c(t)$ than other links during rush hour. For example, at 6 p.m., $P(t)$ is almost $0.6$ for top $0-0.8\%$ EB links, whereas $f_c=0.2702$, indicating that top $0-0.8\%$ EB links are more than twice as likely as all other links to have qualities below $q_c$. Moreover, we shuffled the qualities of top $0-1.6\%$ EB links and found that there is no change in $q_c$ and $f_c$, indicating that congestion of the top EB links still have a strong impact on the network efficiency even if not all of the top EB links are congested. Top $1.6-2.4\%$ EB links show higher $P(t)$ during the morning rush hour but not quite higher $P(t)$ during the evening rush hour. Top $2.4-4.0\%$ EB links show higher $P(t)$ during the evening rush hour but even lower $P(t)$ during the morning rush hour. These results indicate that the highest EB links tend to have lower quality during rush hour compared to links with lower EB, but higher quality during non-rush hour. Fig.~\ref{fig:fig4}(b) and (c) show the $P(t)$ of the intermediate EB links and the lowest EB links, respectively. The lowest EB links tend to have $P(t)$ similar to the average during rush hour, but significantly higher $P(t)$ (i.e., their qualities are relatively low) during non-rush hour. Such a difference in the spatial structure leads to the difference in the global efficiency of the real-world traffic flow network.

Finally, we consider two additional topological properties of links to explore whether there is a topological property better than EB: edge degree~\cite{holme2002attack} and tie range~\cite{park2018strength}. Note that links, edges, and ties are equivalent terms in this article. The degree of an edge connecting node $i$ and $j$ can be defined as: (i) $k_ik_j$ or (ii) $k_i+k_j$, where $k_i$ is the sum of in-degree and out-degree of node $i$. Tie range is defined as the second-shortest path length, i.e., the number of intermediary ties required to reach from a node to its neighbor if their direct tie are removed. We consider these properties because they quantify how much a given link contributes to the network connectivity. Both properties, however, do not perform better than EB in our data. Although this result does not rule out all other possibilities, we think that EB is an effective topological property for our analysis.

\section{Conclusions}\label{sec:concl}
To identify an empirical relation between the network structure and function of urban traffic flows, we analyzed a time-varying real-world traffic flow network and its null-model network with a quality-based percolation method and observed that the real-world network is less efficient than the null model during rush hour, whereas more efficient during non-rush hour. We demonstrate that a difference in the spatial structure of the real-world network during rush hour and non-rush hour is that the quality of the highest EB links is lower than other links during rush hour but higher during non-rush hour, resulting in the difference in the global efficiency.

Our results indicate that urban traffic congestion may not be caused by all traffic flows on the road network being similarly congested, but rather by congestion on parts of the road network that are critical for connecting the whole flow network together. This suggests that these backbone flows and the underlying roads should be considered first when planning urban traffic congestion measures. In the Seoul traffic flow networks, the highest EB links tend to be flows on the highways although there are some exceptions. A recent work also suggests that highways are a key to determine the nature of the percolation transition in urban traffic networks~\cite{zeng2019switch}. Currently, many cities, including Seoul~\cite{SMG, TOPIS}, use citywide or city center average speeds as an indication of how well urban traffic flows are working~\cite{llorca2010traffic, he2016traffic, rahman2022traffic}, but our research shows that this can be misleading because such average speeds overlook the spatial structure of local traffic flows. We believe that our network approach can provide richer information about how urban traffic flows function.

\begin{acknowledgments}
J.-H.J and Y.-H.E. acknowledge financial support by the National Research Foundation of Korea (NRF) grant funded by the Korea government (MSIT) (Grant No. 2020R1G1A1101950). Y.K and Y.-H.E. acknowledge financial support by Basic Science Research Program through the National Research Foundation of Korea (NRF) funded by the Ministry of Education (Grant No. 2018R1A6A1A06024977). All authors thanks to the Seoul metropolitan government for the sharing of Seoul traffic velocity data for this research.
\end{acknowledgments}

\section*{Author Declarations}

\subsection*{Conflict of Interest}

The authors have no conflicts to disclose.

\subsection*{Author Contributions}

{\bf Yungi Kwon:} Conceptualization (equal); Data curation (lead); Formal analysis (lead); Investigation(lead); Methodology (equal); Software (lead); Validation (equal); Visualization (lead); Writing~-~original draft (equal); Writing~-~review and editing (equal). {\bf Jung-Hoon Jung:} Conceptualization (supporting); Data curation (supporting) Formal analysis (supporting); Methodology (supporting); Software (supporting); Validation (equal); Visualization (supporting); Writing~-~review and editing (equal). {\bf Young-Ho Eom:} Conceptualization (equal); Formal analysis (supporting); Funding Acquisition (lead); Methodology (equal); Project administration (lead); Resources (lead); Supervision (lead); Validation (equal); Visualization (supporting); Writing~-~original draft (equal); Writing~-~review and editing (equal).

\section*{Data Availability Statement}
The data that support the findings of this study are available from the Seoul metropolitan government. Restrictions apply to the availability of these data, which were used under license for this study. Data are available from the authors upon reasonable request and with the permission of the Seoul metropolitan government.


\bibliography{reference}

\end{document}